# "PVC Extrusion Development and Production for the NOvA Neutrino Experiment"

R.L.Talaga, J.J. Grudzinski, S. Phan-Budd*
Argonne National Laboratory, Lemont IL 60439 USA

A. Pla-Dalmau, J.E. Fagan, C. Grozis**, K.M. Kephart
Fermi National Laboratory, Batavia IL 60510 USA

**March 19, 2015**

## Abstract

We have produced large and highly-reflective open-cell PVC extrusions for the NOvA neutrino oscillation experiment. The extrusions were sealed, instrumented, assembled into self-supporting detector blocks, and filled with liquid scintillator. Each Far Detector block stands 15.7 m high, is 15.7 m wide and 2.1 m thick. More than 22,000 extrusions were produced with high dimensional tolerance and robust mechanical strength. This paper provides an overview of the NOvA Far Detector, describes the preparation of the custom PVC powder, and the making of the extrusions. Quality control was a key element in the production and is described in detail.

## I. Introduction

**Overview**
The NOvA experiment is designed to search for the appearance, via neutrino oscillation, of electron-neutrinos in Fermilab's NuMI Muon-neutrino beam [REF1, REF 2]. Two liquid scintillator-based detectors, separated by a long baseline, are exposed to a neutrino beam produced at Fermilab. The 300-ton Near Detector is located inside the NuMI beam tunnel, approximately 500 m downstream of the neutrino production target. The Far Detector, located 810 km to the north of Fermilab in Ash River, Minnesota, is substantially larger, with a mass of 14,000 tons to help compensate for the diminished neutrino flux at that distance. Both detectors are located 14 milliradians to the west of the central axis of the NuMI beam in order to intercept a narrower energy range of neutrinos, centered on 2 GeV. The detectors are tracking calorimeters and utilize 16-cell PVC extrusions for the dual purposes of containing the liquid and providing optical segmentation. Approximately 37% of the detector mass is in the PVC structure and 63% of the mass is in the liquid scintillator, both low-z materials, with a characteristic radiation length of 40 cm. This is considerably longer than the radiation length of most tracking-calorimeters, resulting in electron shower lengths in the NOvA detectors that are comparable to Muon track lengths from charged current interactions with NuMI neutrinos.

**Neutrino Detection**
Neutrinos interact primarily with atomic nuclei of the PVC and liquid-scintillator, producing ionizing radiation that excites the liquid scintillator and results in detectable light signals. Each scintillator-filled PVC cell is equipped with a wavelength shifting (WLS) fiber approximately twice as long as the extrusion and looped at the "far end" of the cell (Fig. 1). Generally, the scintillation light is captured by the WLS fiber after several reflections off the cell walls. Simulations show that scintillation light reflects about 8 times on average before entering the fiber. This is the key reason to use highly-reflective PVC surfaces. The Far Detector has a total of 344,064 PVC cells, individually equipped with optical fibers to transport scintillation light to a 32-channel avalanche photodiode (APD) that sits just over an optical connector attached to each 32 cell module.

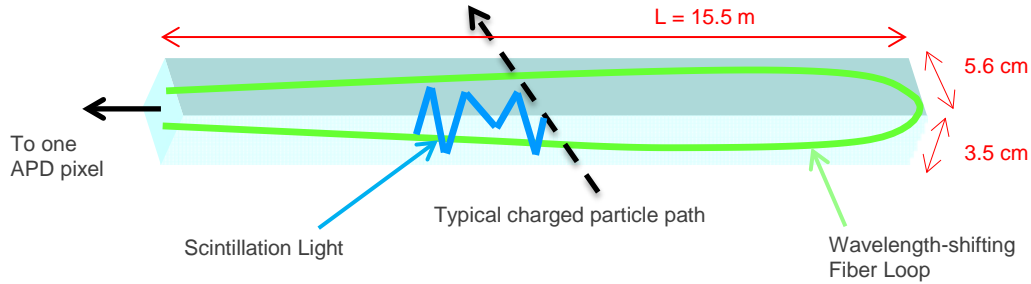

Figure 1: Ionizing particles passing through the scintillating liquid contained within an extrusion cell produce light, which reflects off the PVC walls multiple times until being captured by a wavelength shifting fiber optic loop. Light within the fiber optic travels the length of the extrusion and is detected by an avalanche photodiode (APD). Dimensions refer to *liquid* volume.

**Detector Structure**

The extrusions form the mechanical backbone of the NOvA detectors, providing the strength necessary to maintain a very large structure filled with liquid scintillator. In order to capture the scintillation light for readout, the extrusion cell walls must have a high reflectance; significantly higher than found in commercial PVC products. We have developed a PVC-based formulation to achieve high reflectance while maintaining the necessary mechanical strength. This paper describes the techniques developed to produce more than 11 million pounds of NOvA extrusions that meet strict reflectance, strength and dimensional requirements. An extrusion profile schematic is shown in Fig. 2, and a photograph in Fig. 3a.

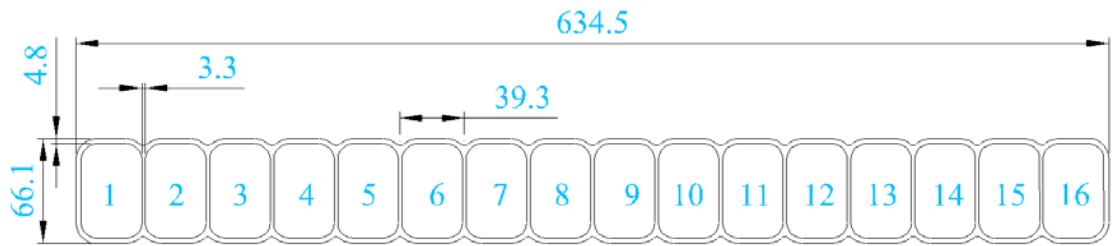

Figure 2: NOvA extrusion cross section with cells numbered 1 through 16 (dimensions are in millimeters).

The Near and Far Detectors consist of free-standing blocks of PVC extrusions, filled with liquid scintillator. The NOvA Far Detector is at this time potentially the largest self-supporting plastic structure ever built. Although the primary purpose of this paper is to describe PVC extrusion development, a brief description of the detector assembly process is helpful to provide context for the physical and optical requirements of PVC extrusions as detector elements.

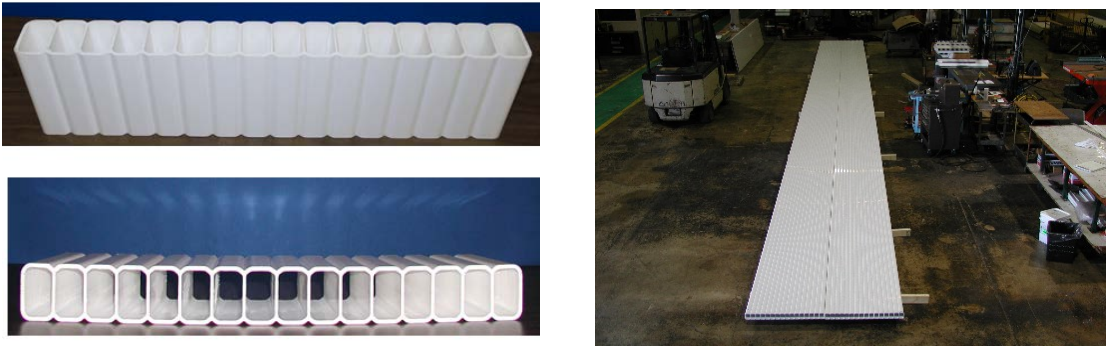

Figure 3: (a) Close-up photos of one 16-cell PVC extrusion, 15 cm long. (b) Two full-size 16-cell extrusions 15.5 m long placed side-by side form the basis for an extrusion module.

After the 16-cell PVC extrusions were produced, they were shipped to the University of Minnesota to be assembled into the basic detector element of NOvA: the extrusion module. A module consists of an instrumented pair of 16-cell extrusions, each 63.5 cm wide and 15.5 m (3.9 m) long for the Far Detector (Near Detector). To make a module, first a pair of extrusions was bonded side-to-side, resulting in a 32-cell object 1.27 m wide as shown in Fig 3b. This was done to maximize efficient use of readout electronics, which is based on 32 channels. Y-11 wavelength shifting fiber (0.7 mm diameter) [REF 3] was inserted down the entire length of each cell and looped around a fixture at the far end, for a total length of approximately 33 meters per cell, depending on the routing distance in the manifold to the optical connector. At the near end, both ends of the fiber were routed inside a manifold to terminate at an optical connector. Both ends of the extrusions were sealed with the aid of custom-made plastic gaskets and adhesive. The near end of a module was sealed with an injection-molded cover that enveloped the fiber manifold and exposed the optical connector. The far end of the module was sealed with a flat PVC plate that was designed to bear a structural load.

The third and final step of the detector assembly process was performed at the experimental sites: Ash River Minnesota (Far Detector) and Fermilab (Near Detector). Both locations required excavation and construction of specialized laboratory detector halls, oriented along the direction of the neutrino beam (approximately a north-south direction). Extrusion modules were shipped from the University of Minnesota to these sites, where they were assembled into detector blocks and placed in position to form the Far and Near Detectors. Because the Far Detector is significantly larger than the Near Detector, PVC extrusions were designed and built to meet the extraordinary size, stress and reflectance criteria necessitated by its requirements. We therefore limit the detector assembly procedure discussion to the Far Detector.

The Far Detector assembly process for each Far Detector Block (FDB) is described in brief. First, twelve extrusion modules were put down next to each other on an assembly platform table. Since each module is 15.5 m long and 1.27 m wide, the twelve modules form a square 15.5 m on a side, with the extruded cells oriented in a north-south direction. A second layer of twelve modules, whose underside was coated with an adhesive [REF 4], was placed on top of the first layer, but with the extrusion cells oriented in an east-west direction, such that the cells of the second layer were orthogonal to those of the first. A third layer of modules was placed and bonded to the second layer, with its orientation in the same direction as the first layer. A fourth layer of modules mimicked the second layer; the fifth layer mimicked the third, and so on, until a total of 32 layers were bonded together. To summarize, each FDB:

- is composed of 384 extrusion modules (768 PVC extrusions)
- consists of 182 metric tons of extruded PVC
- is 2.1 m thick, 15.5 m high and 15.5 m wide

Once assembled, the mechanized assembly table pivoted the FDB around an east-west axis, the proper orientation for use as a detector. Fig. 4 shows (a) the start of the pivoting and moving procedure and (b) the proper orientation of a Block at the Ash River site. After a Block was placed on the detector hall floor, it was filled completely using 700,000 pounds of liquid scintillator. In all there are 28 Far Detector Blocks, consisting of 21,504 extrusions. Near Detector Blocks are made of the same materials but stand only 3.9 m high, one fourth of the height of an FDB.

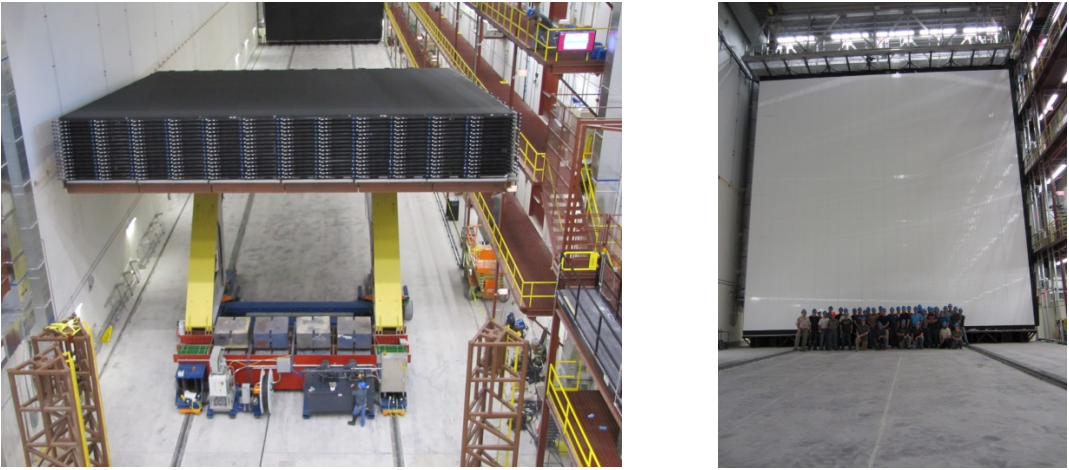

Figure 4: (a) One of 28 Far Detector Blocks is moved into position with the Block Pivoter. (b) A Block is in its final position. The installation crew of 38 people sets the scale perspective. The outside surfaces of Blocks are painted black to prevent external light from passing through the PVC material and being sensed by the APDs.

## II. Reflectivity, Strength and Dimensional Considerations of PVC Extrusions

PVC production consists of two distinct and complex processes, compounding and extruding. A pure PVC resin is compounded with titanium dioxide and several other necessary ingredients to produce a powder that is suitable for use in an extruding machine. The extruding process melts the powder and pushes it through a die to form the desired plastic shape. Because the PVC plastic is very hot at the point of exiting the die, vacuum suction and cooling techniques are implemented to maintain the shape as the plastic cools to room temperature. In each process great care must be taken to preserve the optical and mechanical integrity of the product. In addition to the standard quality control normally used in industrial production, we developed a number of specific tests and instruments to measure output performance and keep losses and waste to a minimum via rapid feedback to the machine operators. The three important qualities tracked or NOvA PVC extrusions were high reflectivity, mechanical strength, and tight tolerances on geometric dimensions.

**Reflectivity**

Maintaining high reflectivity was important in the compounding and the extruding processes. Highly reflective extrusions could not be manufactured if the powder were to be compromised by trace impurities or by improper processing. The extruding process could also compromise reflectivity due to contaminants in the extruding line, extended residence time in the die, or a host of other improper operations. It was important to have an acceptance criterion to quantify the reflectivity of both the powder and the extrusions. Because the scintillation light spectrum is relatively broad, measurement of a single reflectivity in a narrow range of wavelengths was not appropriate. Instead, we devised a method to yield a single numerical value on which the quality could be judged.

The amount of light reflected by an extrusion cell wall is the product of the scintillation light intensity and the reflectance of the PVC cell walls. The scintillation light spectrum peaks around 430 nm with tails extending below 350 nm and above 450 nm, as shown in Fig. 5(a). Because of this broad spectral range and because scintillation light undergoes a large number of reflections before being captured by a WLS fiber, the light-capture process was simulated by a simple Monte Carlo program to quantify the quality of the reflective property of the PVC. The simulation modeled capture of scintillation light in one cell of a NOvA module which was filled with liquid scintillator and instrumented with WLS fiber and readout. The program used wavelength-dependent reflectivity measurements to simulate the number of photons to reach the readout (as a percentage of the light yield of a perfect reflector) to calculate an effective figure of merit called the "light yield".

**Strength**

The extrusions are the primary component of the self-supporting detector structure, leading to specific mechanical property requirements for the PVC. The situation is complicated by the fact that PVC is viscoelastic at room temperature. This implies that the structure will continue to deform over time under a state of constant load (creep). With a design lifetime of 20 years, the most serious implication for the NOVA structure pertains to the structural stability, which is a function of stiffness and the deformed configuration. The continued deformation over time creates the potential that a structure that is initially stable could become unstable in the future, leading to catastrophic failure. Therefore it was important that the creep rate of the PVC be understood and that its value would be sufficiently low. Additionally, since the creep rate is a function of stress, it was important to iterate the extrusion dimensions and overall detector design to minimize stresses and to ensure stability over the long term. This iterative design and analysis required good characterization of the creep properties as well as predictions of the long term behavior. With the custom NOvA formulation, considerable effort was put into this characterization.

The hollow regions of the extrusions are created with metallic inserts within the die which are supported by internal metal structures called spiders. These spiders separate the flow of the PVC melt as it moves through the die, after which the PVC must recombine to form a strong bond.

The interfaces where the material forms a new bond are referred to as "weld" or "knit lines" and are a common feature found in complex extrusions, particularly cellular structures. As the material is extruded, the "knit lines" are created along the length of the extrusion. Typically, knit lines are the weakest part of an extrusion [REF 5]. This is discussed in greater detail in section IV.

**Geometrical Dimensions**

Extrusions made to the specified dimensional tolerances were required for the precision assembly of detector modules and blocks. Important examples were: clearances for the purpose of inserting WLS fibers; fitting endcaps for sealing detector modules; and maintaining flatness for laminating the modules into blocks. Thus the locations and geometry of cell-separators (webs), the radii of curvature of the internal end-cells and flatness of every extrusion had to be within the allowed tolerances. To meet these demands, the PVC compounding had to be kept under tight quality control because inconsistencies in the quality or amounts of additives in the compound as well as unwanted buildup on the die and tooling could easily manifest in the extruding process and bring the extrusions outside geometrical tolerance.

## III. The Making of NOvA-27 Powder

Typical commercial PVC extrusions are made from a "dry blend" or powder, comprised of pure PVC resin, lubricants, stabilizers and other ingredients such as fillers, color additives, and impact modifiers [REF 6]. The "white" PVC formulations normally contain 2 kg to 10 kg of titanium dioxide ($TiO_2$) added to every 100 kg of PVC resin also referred to as 2 to10 per hundred parts of resin (phr) in the plastics industry nomenclature. $TiO_2$ is used to make products white because of its high index of refraction. $TiO_2$ is available in several crystalline structures, most commonly rutile and anatase [REF 7]. The rutile crystalline form of $TiO_2$ ($n = 2.73$) is used in products exposed to sunlight and water, including common PVC applications in outdoor environments. The crystals are coated with inorganic (silica, alumina) and organic (hydrophobic, hydrophilic) treatments for efficient dispersal in the plastics compounding process. The anatase crystalline form of $TiO_2$ ($n = 2.55$) is normally used extensively only in the paper industry. It is untreated and soluble in water and therefore is not stable when exposed to sunlight and water.

Commercial PVC formulations afforded excellent NOvA profiles when tested for the required parameters of mechanical strength and tight tolerances on geometric dimensions. However, their reflectivity was inadequate for NOvA. In consultation with various PVC experts we developed a new formulation that allowed us to optimize reflectivity in a systematic fashion. The grade of PVC and type of $TiO_2$ were selected to provide maximum reflectivity and sufficient mechanical strength. The large and complex NOvA extrusions posed a manufacturing challenge for the PVC powder formulation because of the elevated temperatures and pressures in addition to the residence time and frictional forces in the extruder and die that the material experienced in the extruding process. These conditions required introduction of small but critically important amounts of lubricants, stabilizers and other products [REF 8] to both protect the PVC melt against degradation, and to provide sufficient lubrication to avoid adhesion to the metal surfaces of the extruder and die during transit time. Care was taken to ensure that the additives would not absorb light in the required short-wavelength spectral region.

To maximize the reflectivity and also maintain structural integrity, about thirty custom formulations were compounded and about twenty were extruded for testing. The formulations were made with either anatase or rutile TiO$_2$ in varying amounts. Anatase has a more favorable reflection spectrum above 350 nm, shown in Fig. 5 a, but it is more difficult to compound and extrude than rutile since it is untreated. Tests showed an increase of 14% in the "light yield" figure of merit from extrusions made with anatase over similar extrusions made with an equal amount of rutile (Fig. 5 b). Long-term exposure to sunlight and water were not an issue for this application because the NOvA detectors would be operated indoors.

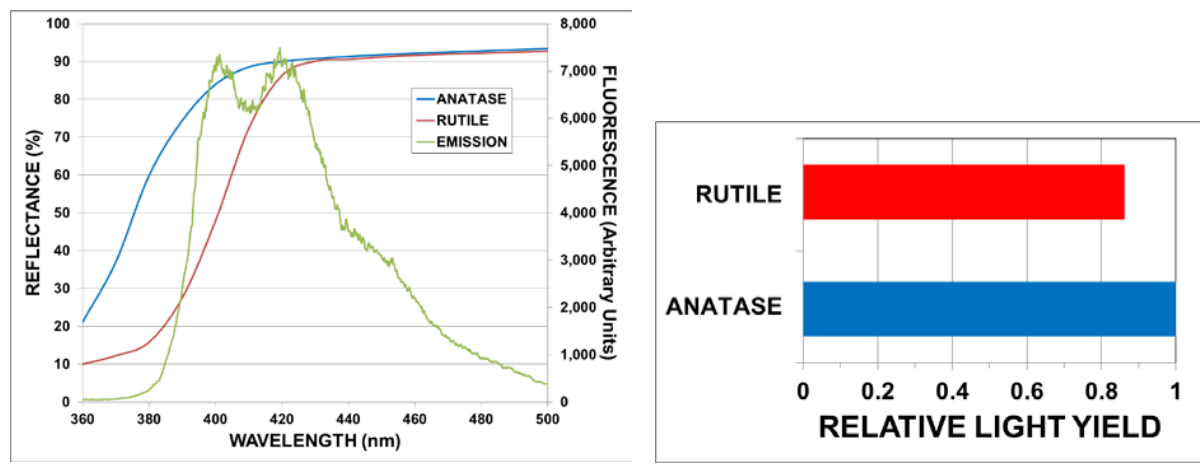

Figure 5: a) Extrusion reflectance spectrum is overlaid with scintillator output spectrum for rutile and anatase TiO$_2$ extrusions. b) The relative amount of light detected for prototype PVC extrusions made with rutile (0.86) and anatase (1.0). The light output was measured in cosmic ray tests.

Reflectivity increased as more TiO$_2$ was added to the compound, up to about 15% TiO$_2$ by weight (19 phr) and stayed constant at higher levels. Besides reflectivity, the mechanical strength and reproducibility of extruded dimensions were optimized. The final formulation for production of NOvA extrusions contained 15% anatase TiO$_2$ along with the ingredients specified in Table 1. This titanium dioxide concentration is twice the highest amounts typically used in industry. The TiO$_2$, along with the other chosen components were added to a pure PVC polymer in a prescribed blending sequence to produce the NOvA-27 (N-27) formulation.

Table 1: NOvA-27 PVC powder formulation.
Note the use of phr (per hundred parts of resin) as the weight units.

| INGREDIENT TYPE | INGREDIENT BRAND NAME | RELATIVE WEIGHT |
|---|---|---|
| PVC | Shintech SE950EG | 100 |
| Tin stabilizer | Rohm & Haas Advastab TM-181 | 2.5 |
| Titanium dioxide (anatase) | Kronos 1000 | 19 |
| Calcium stearate | Ferro 15F | 0.8 |
| Paraffin wax | Honeywell Rheochem 165-010 | 1.1 |
| Oxidized polyethylene | Ferro Petrac 215 | 0.2 |
| Glycerol monostearate | Advalube F1005 | 0.3 |
| Acrylic impact modifier | Arkema Durastrength 200 | 4 |
| Processing aid | Rohm & Haas Paraloid K120N | 1 |
| | Total phr | 129 |
| | wt % titanium dioxide | 15 |

Shintech PVC resin was selected because of its excellent clarity compared to other pure PVC polymers. Advastab TM-182 (20% dimethyl tin, 80% monomethyl tin) was also initially chosen for its higher transparency but it did not stabilize the melt as well as TM-181 (80% dimethyl tin, 20% monomethyl tin) which showed slightly more absorption at low wavelengths. Kronos 1000 was the anatase titanium dioxide that gave the best reflectivity results. The purity of anatase $TiO_2$ became an issue because of the relatively relaxed standards in the titanium dioxide industry, which allowed up to 5% rutile contamination in an anatase product. A 3% or higher rutile concentration in the chosen anatase $TiO_2$ product resulted in an extrusion with lower reflectivity and ultimately lower light yield. An agreement was negotiated with Kronos [REF 9] to hold the rutile concentration to $\leq 2\%$ for the NOvA application. Two different types of impact modifiers were considered (Acrylic and MBS) although not heavily tested since most of the formulations used the same acrylic impact modifier which provided acceptable results. The remaining ingredients provided lubrication at different levels throughout the extrusion process. Several equivalent products were used in the R&D phase with no observable changes in the light yield and mechanical properties of the final extrusion. Once the formulation was settled, product substitutions were not allowed.

Because of the unique characteristics of the custom formulation, including the use and high level of anatase, the extruding company chose not to purchase the compounded PVC powder. Instead, NOvA supplied all of the PVC powder for the extrusions. As a consequence it fell to NOvA to devise methods to maintain PVC powder compounding quality and consistency, while minimizing waste.

PVC powder production did not occur on a dedicated compounding line. Therefore extra care was taken by the selected PVC powder compounding vendor, PolyOne [REF 10], to eliminate contamination from other products and to maintain the N-27 high-reflectance specification. The compounding line was fully scrubbed prior to each N-27 processing. The monthly PVC powder production extended over several consecutive days in one week for about three years. Approximately 200,000 kg of N-27 powder were produced every month. The material was placed into boxes that held approximately 635 kg, resulting in a monthly shipment of about 315 boxes on 11 trucks. The boxes were sent from the PolyOne dry blending facility in Pasadena, TX to Extrutech Plastics in Manitowoc, WI [REF 11] where they were stored on the factory floor prior to use in the extrusion process. A total of 9,484 boxes containing 6,022,340 kg of N-27 PVC powder were processed in the extrusion production phase.

The compounding process was computer-controlled, which is the standard procedure for modern compounding vendors. Holding bins were filled with the ingredients listed in Table 1. The bins were opened according to an optimized timing sequence; the components weighed and released into the blender and mixed by a large blade that produced frictional heat in the process. Alarms were in place that monitored the weights and amounts released into the mix. The process stopped if the quantities were inadequate. A timeline record of the blending sequence and process was provided to NOvA for each production made.

The frequency of Quality Control tests was increased beyond industry norms for the production of N-27 PVC powder. A sample taken from each 635-kg box underwent two tests: (1) the fusion test and (2) the reflectivity test. These results were normalized to a control sample and used to determine the acceptability of the product. Variations in a production batch within the acceptability criteria still caused extruding difficulties, as discussed in the following section.

## IV. The Making of NOvA Extrusions

**Overview**
Extrusions were made on a dedicated production line at Extrutech Plastics, Inc. from January 2011 until December 2013. Unlike the N-27 powder production, the extrusion production line was operated 24 hours per day exclusively for NOvA. Most of the actual production was done in 5 or 6 working days per week, allowing the remaining time for maintenance.

An extruding line has essentially three sections: the extruder and die, the sizing/cooling section, and the pulling and cutting mechanisms. In the first section powder is transported from its container into the extruding machine and turned into a "melt" (Fig. 6a). This is a viscous fluid produced by a combination of external heating elements and friction acting on the powder. The extruding machine [REF 12] chosen by Extrutech was equipped with a custom-made die [REF 13] for the NOvA profile.

Extruding is a push-pull operation. Upon exiting the die, the melt had the shape of the eventual extrusion profile but the material was not yet rigid and therefore subject to collapse (Fig. 6b).

The extrusion's desired shape was maintained in the cooling section until the extrusion was at room temperature. As it reached the last section of the extruding line the extrusion entered the pulling machine, where it was pulled at a speed that matched the throughput of the material coming into the die. This was followed by an automated saw that cut extrusions alternately into two lengths: 15.5 m and 15 cm. The longer length extrusions were destined for NOvA detector modules while the shorter extrusions were used for a battery of quality control tests described in the following sections. These were designed to ensure that the reflective, mechanical and dimensional characteristics of the longer extrusions were acceptable.

**The NOvA Extruding Process**

The extruding machine was equipped with a hopper for powder intake; a heated barrel with a vacuum suction port to remove volatile components; and two counter-rotating screws along the length of the barrel. N-27 PVC powder was loaded into the extruder's hopper in one of two ways, depending on the operational circumstances. During normal operation, the powder was loaded from a "day bin", a large container that held the blended contents of three 635-kg shipping boxes to reduce possible box-to-box variations. At other times, when necessary, the powder was loaded directly into the hopper from a single box. Gravimetrically measured quantities of powder were delivered into the barrel at a rate of approximately 390 kg per hour. Four zones of external heating elements lined the barrel. The mixing screws were 2.88 m long with a length-to-diameter ratio of 32:1 and were modified for our application to provide additional frictional heating and to increase the mechanical torque. At the die, the melt temperature was approximately 188 °C (370 °F) and the melt pressure was 180 bar (2,600 psi).

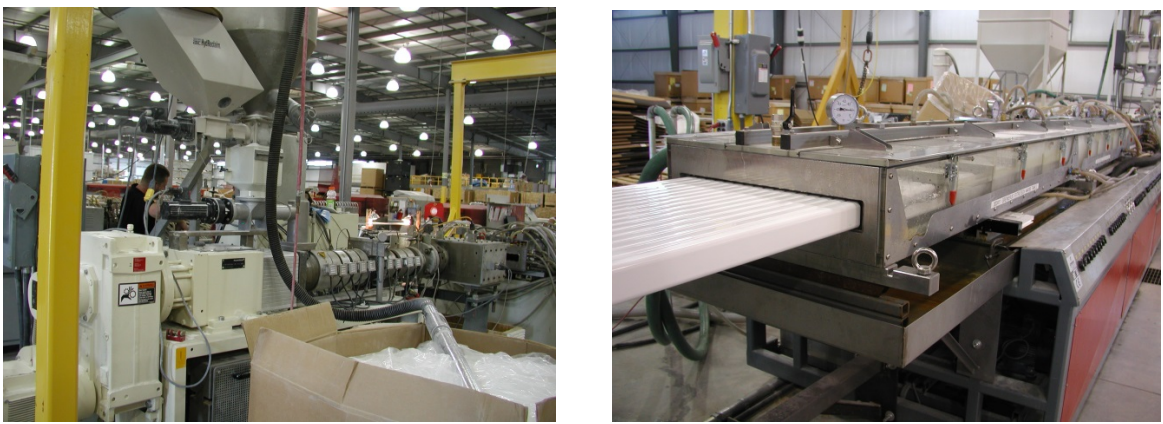

Figure 6: (a) N-27 powder is transported from a plastic-lined "gaylord" container into the extruder's hopper (top-center of the photo). The extruding barrel and back-side of the NOvA die (right-center of photo) are also visible. (b) Rigid extrusion exits the water cooling tanks and is about to enter the puller.

As the melt flowed through the die, it was broadened from the 9.0 cm diameter barrel to the 64.1 cm width of the die and also separated into a number of parallel directions to produce an extrusion with 16 cells. As it traveled through the die compartments, the melt was recombined at various stages, both transverse and longitudinal to the flow, resulting in a final recombination at 77 different locations across the extrusion profile. Knit locations for a typical cell are shown in Fig. 7. Extensive development of the N-27 compounding, die design, and extruding processes led to extrusions with strong knit lines and high intrinsic plastic strength, acceptable for assembly into the NOvA detectors. The strength of the base material is quantified in the next section. More information regarding the knits in NOvA extrusions is provided in [REF 14].

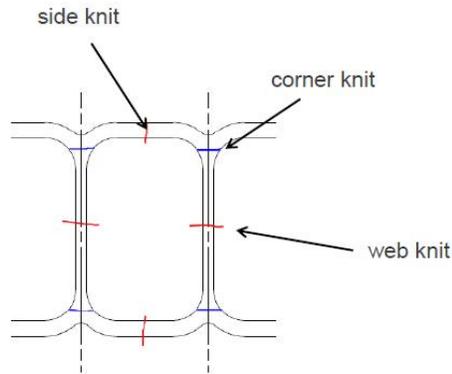

Figure 7: Locations where knits are formed for cells 2 through 15.

Upon exiting the die, the newly-formed extrusion was pulled through two sizing tools, called calibrators, to cool the PVC and to retain the extrusion's shape. The tunnel-like calibrators were water-cooled and had numerous pinhole vacuum ports to draw the hot extrusion tight to their inner surfaces, maintaining shape as the extrusion passed through and into water cooling tanks. The tanks were sealed and held under partial vacuum to keep the warm extrusion from sagging. The amount of cooling for NOvA extrusions was greater as compared to normal commercial extrusions because of the thick walls and high throughput of PVC material.

After exiting the last water-cooling tank, the extrusion was sufficiently rigid to enter the pulling machine. Tractor-like treads captured the extrusion from above and below to achieve sufficient frictional grip and pull the extrusion at a constant speed as show in Fig. 8.

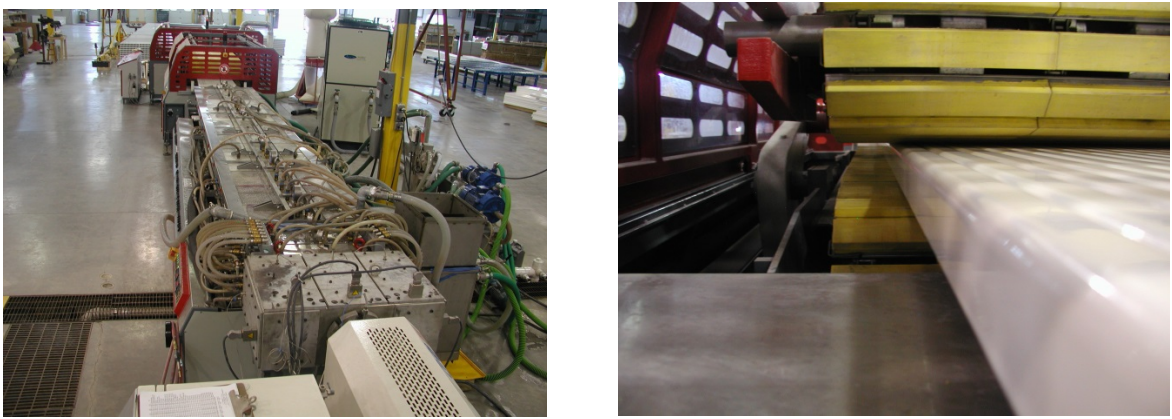

Figure 8: (a) Extrusion line. Shown in sequence are: the end of the extruding barrel (bottom of photo), the die, the sizing/cooling section, and the puller (red). (b) Close-up photo of the puller showing treads pressing down on and pulling an extrusion.

The final unit in the extruding line was the traveling automated saw. When an extrusion was pulled 15.5 m (15 cm) beyond the saw, it was cut and moved from the line. The saw blade moved longitudinally with the same speed as the extrusion as it cut, ensuring a perpendicular cut with minimal binding of the saw blade.

**Extruding Problems and Solutions**

The tight geometric, reflectance and mechanical tolerances, coupled with a custom PVC powder formulation resulted in a narrow operating window. Occasionally extrusions did not meet the specifications. To correct problems, the first response was to adjust operating parameters. Because changes to extruding parameters took on the order of an hour to be realized in the extruded product, it was likely that at least several hours would be needed to determine if the problem was with the extruding process or with the powder. This loss of time and money was not acceptable. We developed a method to determine the source of the problem quickly by exchanging the boxes from the currently-designated lot of powder with boxes saved from an earlier lot, which was known to contain already proven powder from prior extruding experience. If the extrusions did not meet specifications soon after the reserved powder was introduced, the fault was associated with the extruder. The extruding machine was shut down and the die and other suspected components were disassembled and cleaned.

If, after the reserve powder was introduced, the extrusions did meet the specifications, the boxes containing the current powder lot were set aside. In some cases such boxes were returned to the powder manufacturer and replaced with a new lot at no charge. In other cases, an on-site blending procedure combining powder from acceptable lots with the suspected lot in the "day-bin" resulted in acceptable extrusions.

Generally, strength and reflectance properties of extruded material did not change quickly (over the time span of one or two hours). If they did, the change would usually be noticed by the extrusion machine operators as a change in pressure or machine screw torque and adjustments would be made.

**Production**

Approximately forty of the 15.5 m long extrusions and accompanying 15 cm samples were made in one 24-hour period, one pair every 35 minutes. This continuous 24-hour operation required a crew of four Extrutech technicians for every 8-hour shift. Quality control was of the essence because of the stringent physical and dimensional constraints. We developed a rigorous quality control protocol, significantly beyond the industry-standard methods employed at commercial PVC extruding plants. A dedicated on-site NOvA technician was stationed on a full-time basis at the Extrutech plant and was responsible for updating and maintaining the flow of data to the database and for proper implementation of the quality control tests. Instructions on the use of quality control instruments and procedures were given to Extrutech staff, who executed them. Quality control data was used for immediate feedback to machine operators if dimensional tolerances were close to being violated. Quality data logs were uploaded to the NOvA hardware database on a daily basis and information on the entire production run was used to detect deviations from specifications. A user interface program accessed the database and made histograms and time-trend plots of a variety of measurement, available for viewing with a web browser. A description of the quality control instruments and procedures is provided in Section V.

After initial difficulties in the transition from R&D extruding to mass production, the amount of unusable material decreased and the production became efficient. The scrap rate, defined as the weight of PVC used for acceptable extrusions divided by the total weight of extruded powder was approximately 8%, a reasonable value in light of the fact that NOvA extrusions were made with a unique PVC powder.

**Handling, Transport & Storage**

The NOvA 15.5 m-long extrusions weigh approximately 236 kg and needed to be picked up, moved and placed into a storage stack by mechanical means. This was accomplished by a vacuum lifting device attached to a gantry hoist. Suspended in this way, the load was moved to various locations on the factory floor until it was stacked for shipping, as shown in Fig. 9.

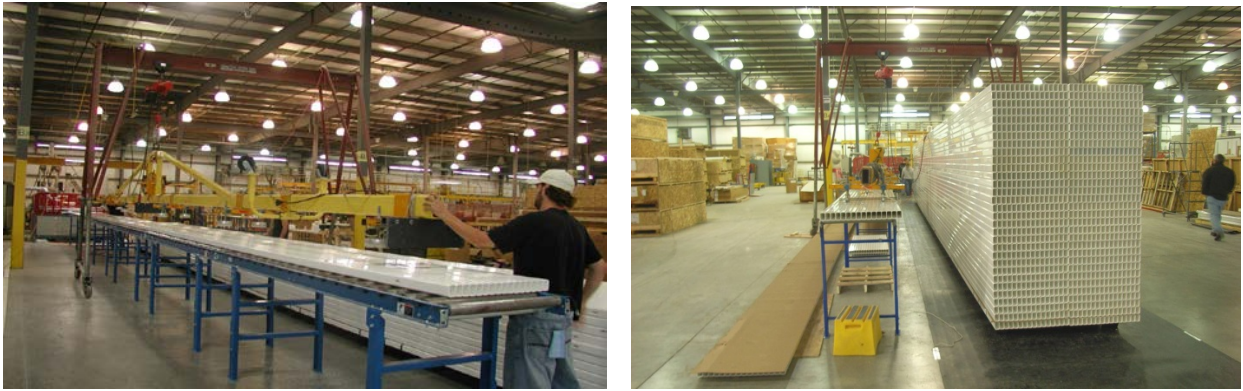

Figure 9: (a) The vacuum lifter, shown just above the extrusion, is supported by a movable gantry. An electronic weight scale is attached just below the hoist, above the yellow frame. (b) The vacuum lifter is used to stack extrusions for transport to the module asembly factory.

Requirements for handling, shipping and storage of extrusions fell into three categories: safety of personnel; affordability of transport of the extrusions; and structural integrity of the extrusions while handling, shipping and storing. NOvA performance parameters dictated a high degree of flatness for each extrusion both during the active lifting phases and the passive storage phases of production. It was relatively easy to converge on processes that satisfied all necessary conditions.

The vacuum lifter system picked each extrusion off the end roller table of the extruder. Figure 10 shows the custom cups [REF 15], molded of a soft silicone so they could contour the scalloped profile of the extrusion surface easily. The number and spacing of cups was determined by the weight of extrusions and the amount of allowable deflection.

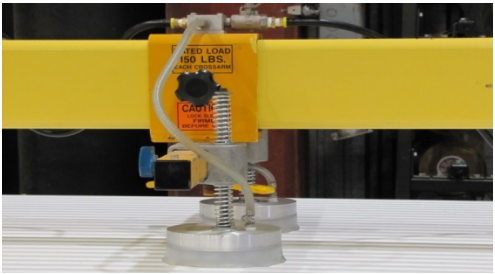 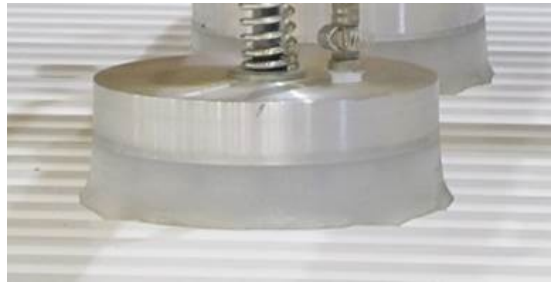

Figure 10: (a) Custom-made silicone cups used to lift the extrusion. (b) Close-up photo of the silicone cup conforming to the scalloped surface of an extrusion.

A support base for the extrusion stacks was devised using commercially available plastic pallets and a layer of extra ablative extrusions. Eight pallets were affixed to the pair of base extrusions utilizing polyester strapping. The seven spaces between the distributed eight pallets were of a size to allow additional pallets to sit between the fixed ones during storage periods. This provided a uniform platform to ensure minimal deflection during storage while leaving spaces for the hydraulic jacking system when moving the load onto and off of a truck and around factory floors. A finite element analysis was performed [REF 16] to validate the design (Fig. 11).

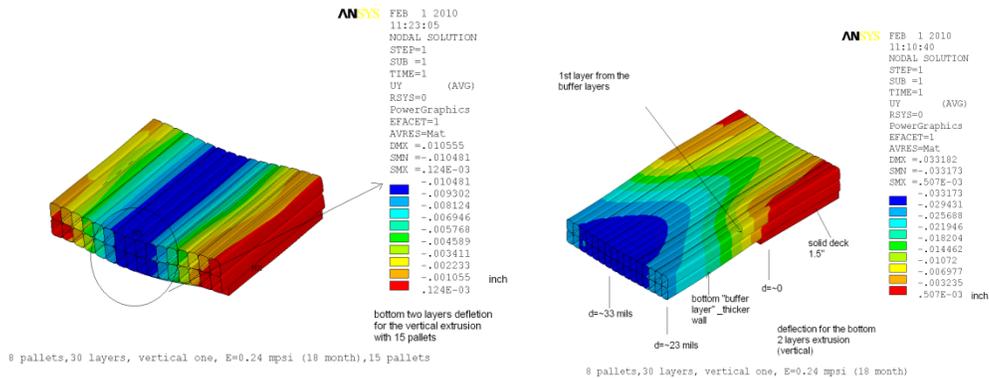

Figure 11: (a) Deflection of bottom two layers supported by 8 pallets. (b) Deflection of bottom two layers supported by 15 pallets.

Once stacked, a load of extrusions had to be moveable. For stability the stacks were strapped with commercial ratchet straps. Initially a plan was devised to use an air caster or float pad system. Unfortunately the floors could not have any cracks or debris on them and truck beds needed a cost-prohibitive liner for the floors. The hoses and control boxes required for this system were unwieldy and made moving around the factory difficult. Slopes into the truck made steering dangerous for personnel and equipment in the area. Ultimately the air caster system was replaced by a series of hydraulic jacking platforms (Fig. 12).

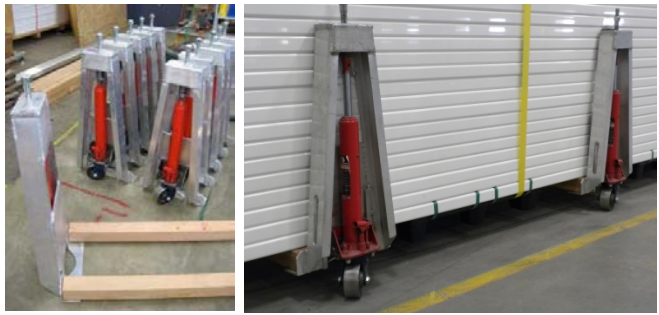

Figure 12: (a) Hydraulic caster jacks. (b) Jacks inserted under a stack of extrusions.

The NOvA production extrusions fit into a standard USA 53-foot dry van trailer. One trailer could contain a stack two extrusions wide by 28 high without special accommodation for loading/unloading and while keeping the load under the standard USA 40,000 pound weight limit. A strapped and jacked load was steered into a dry van using a small forklift (Fig. 13a). Once inside the load was stabilized with standard commercially available cardboard bulkheads and inflatable air bags (Fig. 13b). A one-time use temperature indicator was attached to the load to monitor high temperatures in the summer due to the possibility of deformation of the extrusion walls, especially in the strap region. Transporting during cooler times of the day was sufficient to avoid deforming the extrusions. Removal from the truck van merely required deflating the air bags, replacing the jacks and pulling the load with a forklift.

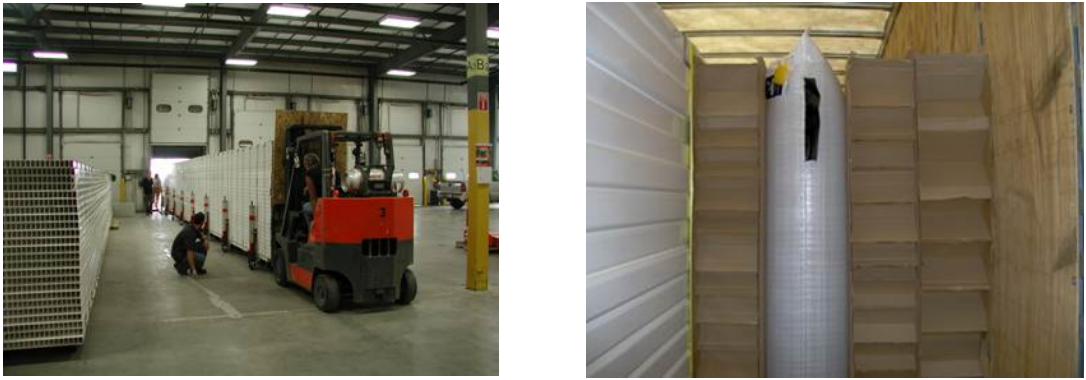

Figure 13: (a) A forklift was used to push one extrusion stack (2 x 28 extrusions) into a standard "53-foot" covered trailer. (b) The stack was braced for shipment with corrugated cardboard and inflatable airbags.

## V. Extrusion Quality and Control

A series of real-time quality control (QC) measurements was performed on extrusions continuously throughout the production period and recorded to the NOvA data base. The starting point was a visual inspection of the long extrusion, which was then weighed and moved to the vacuum-test station. The extrusion weight was monitored to control PVC cost, determine the detector mass, and as an indicator of extruder performance. Dimensional tolerance allowed weight variation, which was minimized to conserve material. The average weight of extrusions was 236.5 kg with a standard deviation of 3.9 kg.

After the visual inspection the accompanying 15 cm extrusion was placed on an automated optical scanner for a precise measurement of its cross sectional profile. Then the 15 cm extrusion was subjected to one of two mechanical strength measurements: either the drop-dart measurement or the hydraulic pressure test measurement, in alternating order. After the strength tests several disks were cut from unaffected webs to measure the reflectivity of the inner surfaces. The overall sequence of this QC process is illustrated in Fig. 14.

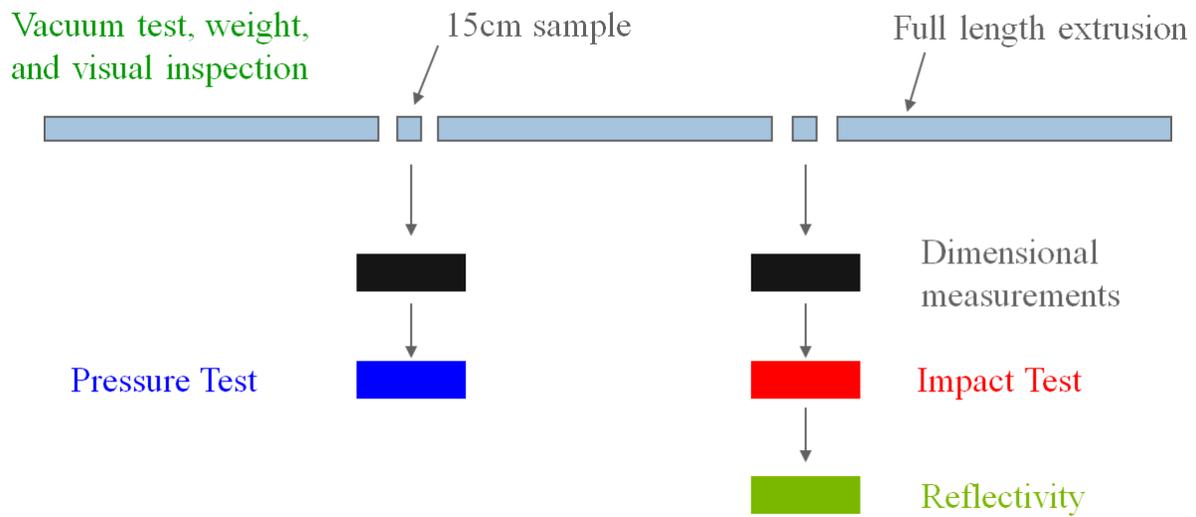

Figure 14: The sequence of extruding QC tests, with extrusions moving from right to left, is shown. Refer to the text for details.

**Visual Inspection**

Visual inspections were used to detect trends in the process that would eventually lead to unacceptable production extrusions. Extrusions were inspected for surface quality, which could indicate slight changes in the composition and quality of the powder or extruding parameters. Discoloration or streak lines would indicate residue buildup in the die; a warning that the die required cleaning.

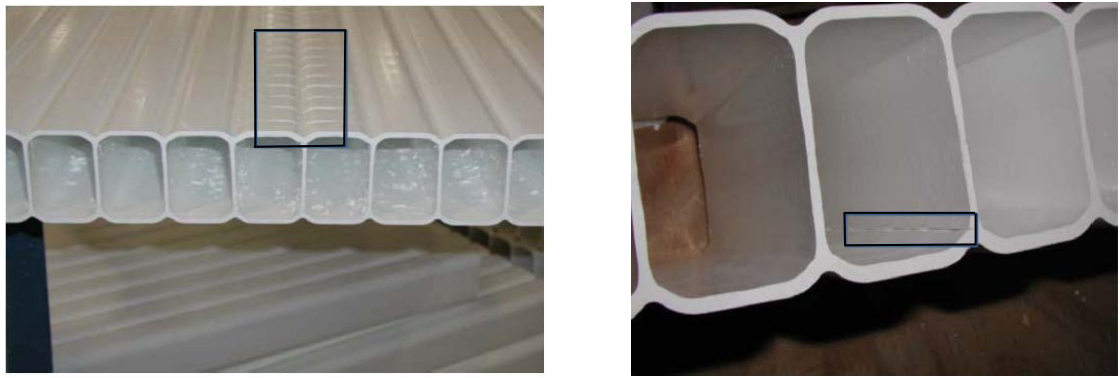

Figure 15: Example of extruding problems detected by visual inspection. Rectangular outlines added to the photos enclose (a) chatter marks on the top surface (b) a poor knit in lower right corner.

**Vacuum Test**

After the visual test, the long extrusion was moved from the extrusion line to the vacuum test station. The vacuum test machine used two easily and quickly applied end seals (Fig. 16) and was PC controlled. The test was designed to detect defective extrusions by observing the rate of vacuum decay. The test routine consisted of three independent tests: Test T1 evacuated cells 2 through 16 (even-numbered cells), Test T2 evacuated all cells (cells 1-16), and Test T3 evacuated odd-numbered cells (1-15). An automated program operated the vacuum pump and valves, recorded the pressures throughout the vacuum test, and displayed the test results. The vacuum test alternately subjected the even and odd-numbered cells to vacuum. This resulted in an atmospheric pressure load on the outer walls of the cells under vacuum. Similarly, the internal webs were subjected to the same pressure acting toward the cell interior. Fig. 17 illustrates the resulting forces as they would act on the internal cells (2-15). This test provided a check of the web integrity over the whole extrusion without being destructive and was complementary to the hydraulic test, described later. A failed vacuum test would have indicated a leak in an internal web or an external wall that would otherwise have been difficult to detect by visual inspection. There were no failures detected during the entire extrusion production.

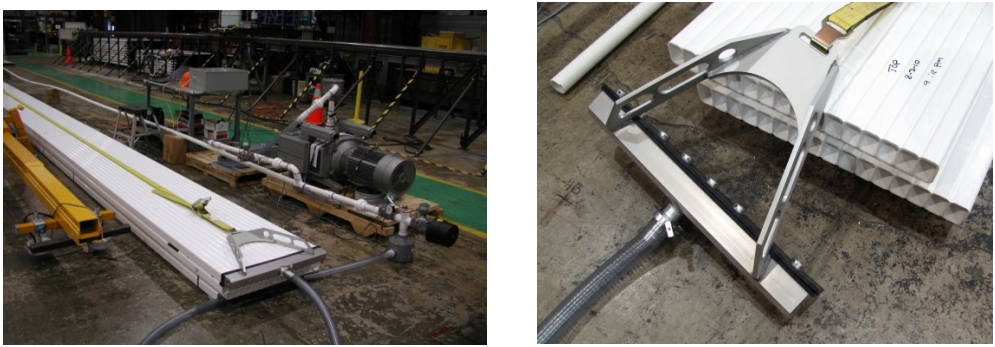

Figure 16: (a) The vacuum tester, with end-sealing manifold in place and the vacuum pump. (b) Close-up of the manifold, showing vacuum ports on odd-numbered cells. A second manifold on the other end evacuated the even-numbered cells.

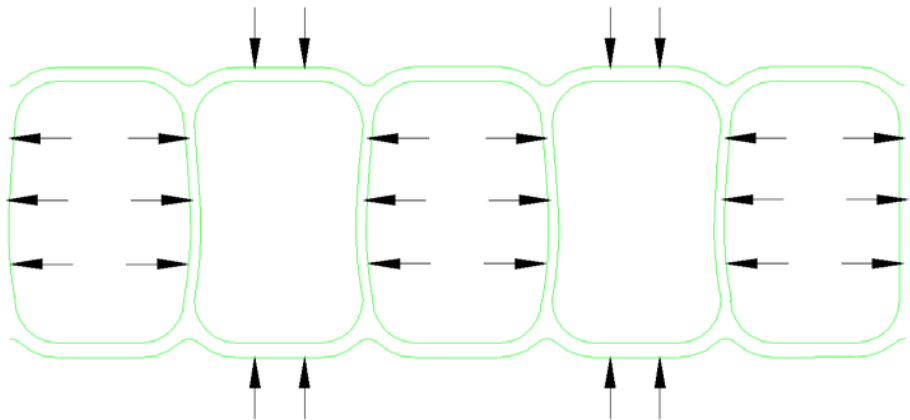

Figure 17: A schematic of the vacuum test, illustrating the resulting forces as alternate cells are subject to vacuum. The figure only shows five of the 14 internal cells.

**Dimensional Measurements**
The tight tolerances on the geometrical dimensions necessitated a thorough measurement of each 15 cm extrusion. Extrusions are not machined objects. The dimensions will vary from one **extrusion to another in the short term and over the course of production. A commercial optical** metrology machine [REF 17] with custom scanning and analysis algorithms was used for this purpose. The duration of each scan was approximately 20 minutes, shorter than the 35 minutes required to produce a long extrusion. Over 150 dimensions were checked using the optical measuring device. All acceptable extrusions were required to meet the key dimensions and tolerances listed in Table 2.

Table 2: Key dimensions used for acceptance of extrusions.

| Dimension | Key Tolerance | Feature Tolerance [mm] (min/max) | As Built (average +/- standard deviation) |
|---|---|---|---|
| Flatness | max | 0.0 / 0.5 | 0.33±0.08 |
| Web minimum thickness | min | 2.9/4.0 | 3.3±0.2 |
| Web location stack-up | min/max | ±2.0 | ±0.28 |
| Web perpendicular | max | 1.0 | 0.4±0.15 |
| Top wall minimum | min | 4.5 / 5.5 | 5.0±0.16 |
| Bottom wall minimum | min | | 4.9±0.1 |
| Outer wall min thickness | min | | 4.9±0.16 (1) 4.9±0.14 (16) |
| Individual cell height (edge to edge) | min/max | 66.1 ± 0.5 | 66.2±0.1 |
| Individual cell height (from datum) | max | 66.1 ± 0.5 | 66.1±0.1 |
| Trough | min/max | 56.5 / 59.5 | 57.31± 0.15 |
| Internal radii, outside corners cells 1,16 | min/max | 8.0 ±0.5 | 8.2±0.19, 7.6±0.14 8.0±0.17, 7.7±0.16 |
| Extrusion Width | min/max | 634.5 ± 1.0 | 634.55 ± 0.25 |

Precise measurements included the minimum thickness of the webs, inner radii of curvature of cells 1 and 16, thickness of top and bottom walls, as well as the overall flatness across all 16 cells. The metrology machine and 15 cm sample are shown in Fig. 18.

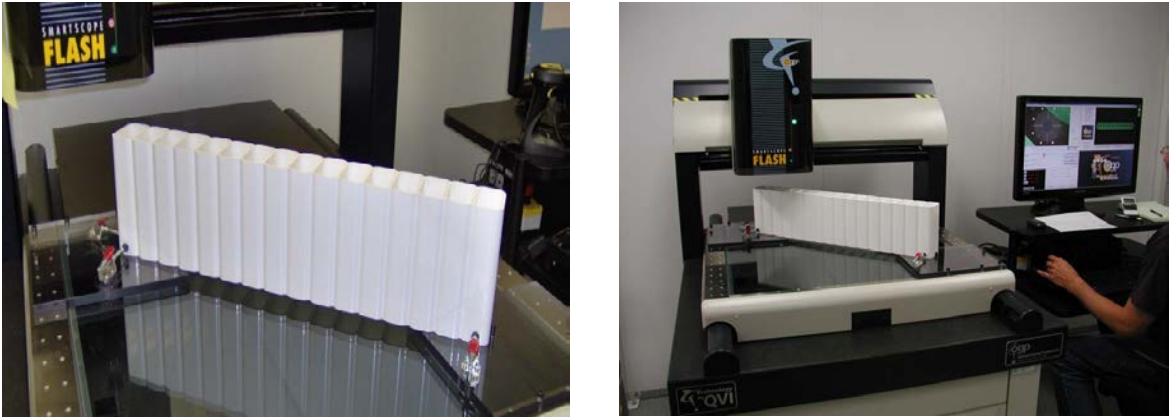

Figure 18: (a) A 15-cm extrusion placed on the scanning bed of the metrology machine. (b) The machine and operator console.

**Impact Test**

A drop-dart impact test was performed on alternating 6 inch samples. Fixturing is required for repeatability of the drop dart test, which is dependent upon strike position. Due to the quantity of tests and the time involved, the standard drop dart test was configured into an automated machine using a pendulum to deliver the dart impact. The custom machine and an impacted cell are shown in Fig. 19. As a consequence of the scatter in this test, the data was compared to a minimal threshold value and to a historical trend rather than to a strict reject criterion. When the impact values were lower than the threshold this generally served as an indicator that an adjustment in the extrusion process parameters or substitution of the PVC powder lot was required. In order to attempt to save extrusions associated with failed drop dart test, tensile testing was performed to determine if the extrusions maintained sufficient ductility to be acceptable.

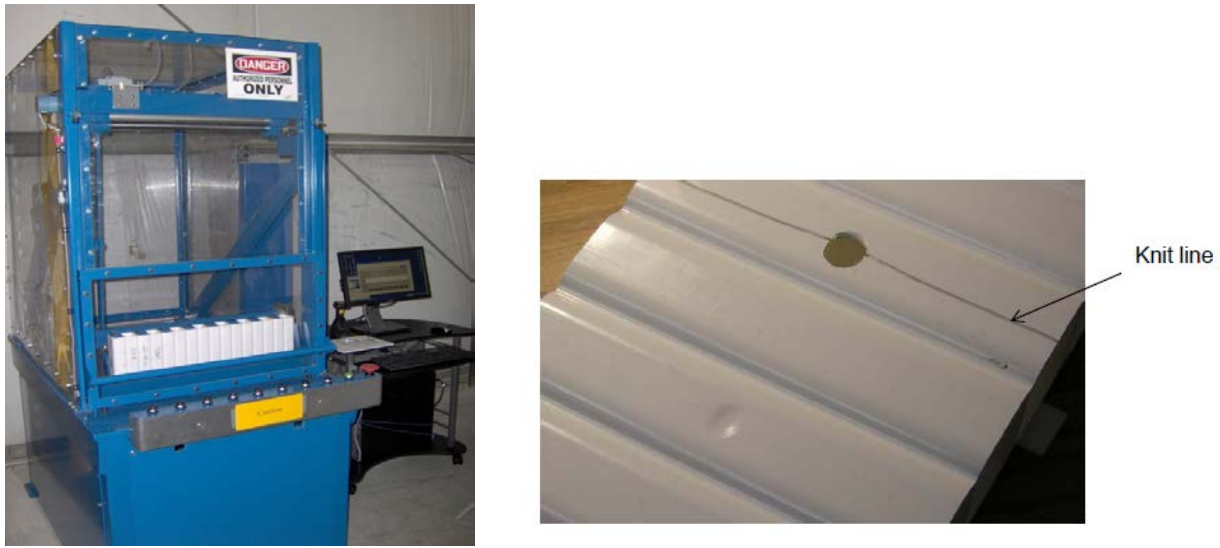

Figure 19: (a) The customized automated drop dart impact machine and (b) the resulting impacts on cells. Note the crack along a knit line in the upper cell.

**Hydraulic Test**

A hydraulic pressure test was performed on the 15 cm extrusion not subjected to the drop dart impact test. Since the pressure was increased until failure occurred, the test was intentionally destructive. While the highest nominal pressures in the detector were less than 1.4 bar (20 psi), attaining a pressure of 10.3 bar (150 psi) was used as the passing criteria for the test. The pressure to fail was routinely well over 13.8 bar (200 psi).

The test was performed by clamping a sample at each end, filling alternate cells with water and then increasing pressure, illustrated in Fig. 20. The webs were subjected to bending and tensile stresses due to the applied pressure. The bending stress was intended to expose poor knits. The pressure test was alternated on subsequent extrusion samples such that the odd cells would be pressurized in one test followed by the even cells in the next test.

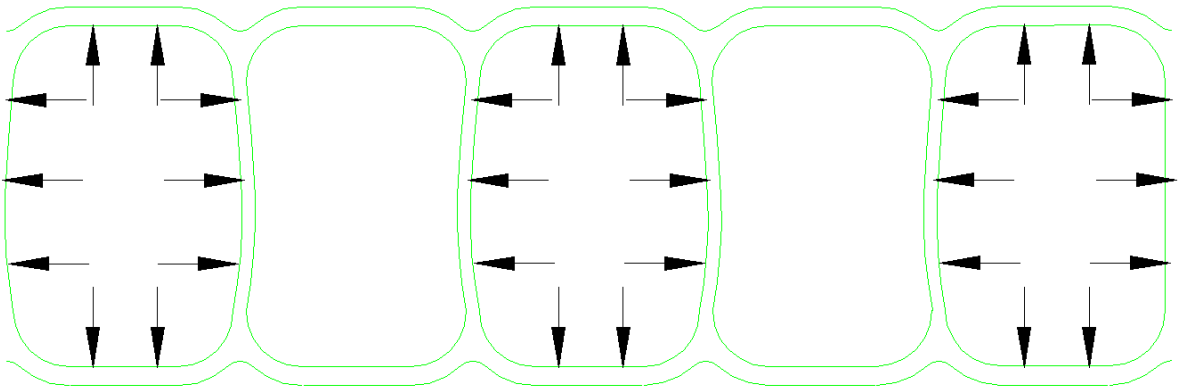

Figure 20: A schematic of the forces due to the hydraulic pressure applied to alternate cells.

**Tensile Test**

Special extrusion samples were cut once per day at various times and sent to ANL. Samples were shipped weekly such that there was about a two week lag between production and testing. As noted above, the impact test served as the real-time check of the mechanical properties. In this regard the tensile test was an assurance check. This tensile test followed ASTM D638. The desired minimum properties for NOvA are listed in Table 3. The typical stress-strain response from the tensile test is shown in Fig. 21. In general, the PVC was quite ductile. Fig. 21 shows the ultimate tensile strength (UTS) and total elongation of a large fraction of the specimens. The UTS is quite consistent and the elongation is generally above 15% and mostly attaining 50%.

Table 3: Minimum mechanical properties desired in tensile testing.

| Material Property | Value | Test Method |
|---|---|---|
| Modulus of Elasticity (Instantaneous, t=0) | 450,000 psi | ASTM D638 |
| 0.2 % Offset Yield | 4000 psi | ASTM D638 |
| Ultimate tensile Stress | 5500 psi | ASTM D638 |

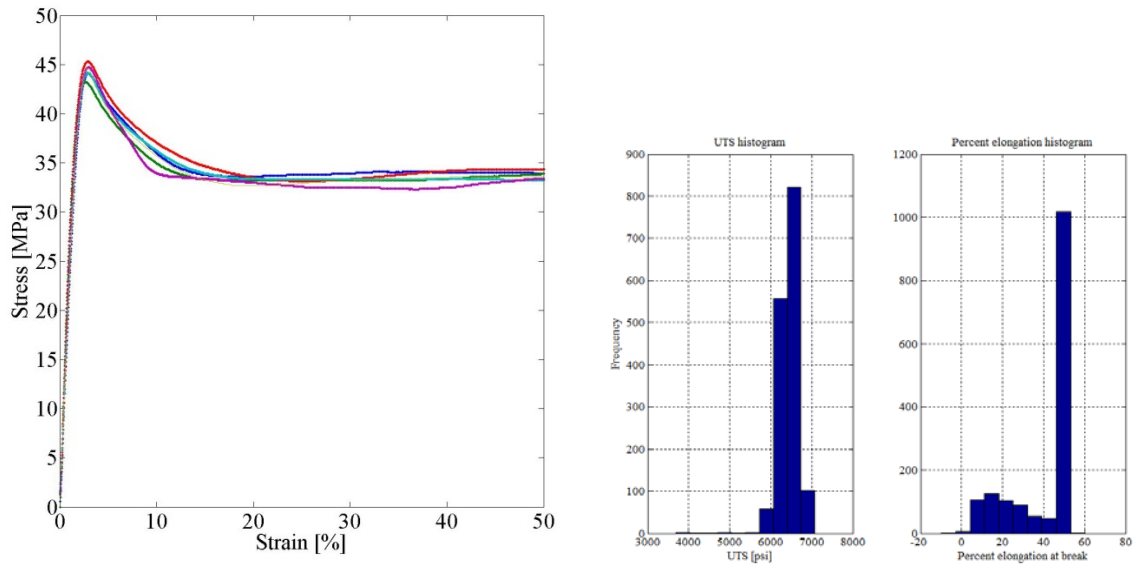

Figure 21: (a) Typical stress-strain response from a tensile test (stopped at 50% elongation). Histogram (b) shows the ultimate tensile strength (UTS) and histogram (c) shows the percent elongation.

**Reflectivity Test and Light Yield**
The reflectivity data were recorded for the wavelength range from 360 nm to 500 nm using a spectrophotometer [REF 18]. The measurement used 4-cm diameter discs cut from web number 15 of each even-numbered QC sample after the impact test was completed. Three discs from the web were separately checked and their results averaged by the instrument. In order to provide a pass-fail decision, the measured reflectivity spectrum was incorporated into a Monte Carlo simulation. This simulation was calibrated to actual light yield data available from a NOvA prototype detector. Tests with the prototype demonstrated a suitable light output for NOvA physics analysis needs [REF 1], with a minimum acceptable light yield value of 14.6% as referenced to a perfect reflector. Extrusion samples with light yields equal to or above 14.6% passed the reflectivity test. Approximately every second extrusion sample was tested in this manner. The few extrusions below this limit were put aside and further tested to gather information on the nature of the problem causing the decrease in reflectivity. The light yield is shown in Fig. 22 for all measured extrusions used in the construction of NOvA. The average value of the light yield for those extrusions is 16.9% ±1.1%. The average light yield as a function of number of produced extrusions is shown in Fig. 23. The light yield was relatively stable throughout the production period, and became more stable as time progressed.

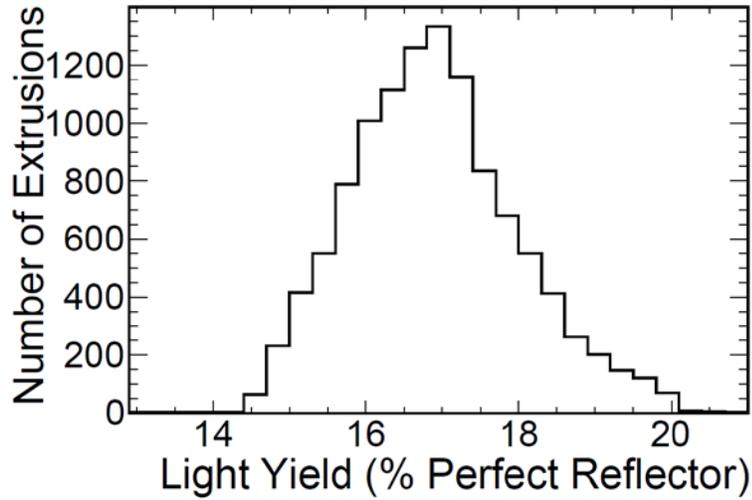

Figure 22: The histogram of light yield values from all measured and accepted extrusions.

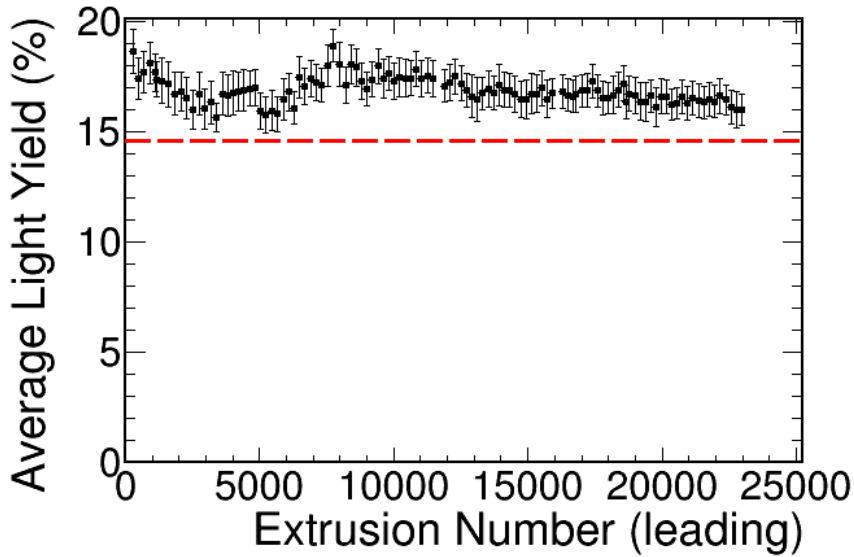

Figure 23: The average light yield as a function of extrusion number from measured and accepted extrusions. The results of the light yield test are averaged for 100 consecutive measurements. This average is plotted as a function of the extrusion number of the first measurement, with the error given by the standard deviation.

## VI. Summary

Large PVC extrusions with high dimensional tolerance and robust mechanical strength as well as excellent reflectivity were produced for the NOvA neutrino oscillation experiment. Production was based on extensive R&D activity over several years. After initial tests with smaller existing dies, it became apparent that a full-size die was required to optimize the PVC compound and the extruding technique. Based on the R&D experience, new extrusion tooling was acquired and commissioned in 2010 for production, which began in January 2011. Initially the full-scale production process did expose some difficulties as we transitioned from R&D to production. Several causes were identified, which included powder consistency, the initial die design and the extruder operation parameters. After the problems were resolved, typical scrap rates were approximately 8%, a value deemed reasonable in light of the fact that NOvA extrusions were made with a unique PVC powder. A robust QC / QA program was instituted early and provided the framework for identifying parameter trending in addition to qualifying suitable production extrusions.

Ultimately the total number of acceptable extrusions produced and shipped for module assembly was **22,190**, enough to produce the Far Detector including spares as well as extrusions used in the construction of the Near Detector.

**Acknowledgements**

We would like to acknowledge and thank M. Klauck, J. VanGemert, G. Gillespie, K. Wood and Z. Matjas for their assistance in extrusion production, database implementation and quality control testing; H. Zhao and R. Kellogg for their contributions to fabricate and automate quality control instrumentation; R. Fischer for studies to improve the structural integrity of extrusions; J. Summers for guidance in developing the PVC formulation; G. Sheehy and the staff at Extrutech Plastics, Inc. for their guidance and support over many years; D. Piel, B. Baran and the staff at PolyOne Corporation for their input in PVC compounding; The University of Minnesota's module assembly factory staff (Department of Physics and Department of Mechanical Engineering) for their valuable feedback throughout the extrusion production effort.

Argonne National Laboratory is operated by UChicago Argonne, LLC under contract No. DE-AC02-06CH11357 with the U.S. Department of Energy, Office of Science. Fermi National Accelerator Laboratory (Fermilab) is operated by Fermi Research Alliance, LLC under Contract No. DE-AC02-07CH11359 with the U.S. Department of Energy, Office of Science.

* Present address: Physics Department, Winona State University, Winona, MN USA
** Present address: Extrutech Plastics, Inc., Manitowoc WI